\documentclass[twoside,leqno,twocolumn]{article}

\usepackage{xcolor}

\definecolor{blue-vlight}  {rgb}{0.871, 0.937, 0.988}
\definecolor{red-vlight}   {rgb}{0.984, 0.871, 0.878}
\definecolor{myblue}         {rgb}{0.337, 0.592, 0.773}
\definecolor{myred}          {rgb}{0.776, 0.357, 0.396}

\usepackage[letterpaper]{geometry}
\usepackage[
colorlinks=true,
linkcolor=myred,
urlcolor=myblue,
citecolor=myblue,
linkbordercolor=blue-vlight,
urlbordercolor=blue-vlight,
pdftex
]{hyperref}
 
\usepackage[ruled,noend]{algorithm2e}
\usepackage{subcaption}
\usepackage{stfloats}
\usepackage{ltexpprt}
\usepackage{booktabs}
\usepackage{amssymb}
\usepackage{wrapfig}
\usepackage{balance}

\usepackage{ifthen}
\usepackage{xspace}
\usepackage{amsmath}
\usepackage[disable]{todonotes}

\usepackage{xspace}
\newcommand*{\eg}{e.g.\@\xspace}
\newcommand*{\ie}{i.e.\@\xspace}

\newcommand{\Psrdm}[1]{\ifthenelse{\equal{#1}{*}}%
  {\textsc{SRDM}\xspace}%
  {\textsc{Scheduling with Release Times and Deadlines on a Minimum Number of Machines}\xspace#1}%
}
\newcommand{\Ptcpsp}[1]{\ifthenelse{\equal{#1}{*}}%
	{\textsc{TCPSP}\xspace}%
	{\textsc{Time-Constrained Project Scheduling Problem}\xspace#1}%
}
\newcommand{\Prcpsp}[1]{\ifthenelse{\equal{#1}{*}}%
	{\textsc{RCPSP}\xspace}%
	{\textsc{Resource-Constrained Project Scheduling Problem}\xspace#1}%
}
\newcommand{\Pfpsp}[1]{\ifthenelse{\equal{#1}{*}}%
	{\textsc{FPSP}\xspace}%
	{\textsc{Flexibilization Project Scheduling Problem}\xspace#1}%
}
\newcommand{\Pracp}[1]{\ifthenelse{\equal{#1}{*}}%
	{\textsc{RACP}\xspace}%
	{\textsc{Resource Acquirement Cost Problem}\xspace#1}%
}
\newcommand{\Psracp}[1]{\ifthenelse{\equal{#1}{*}}%
	{\textsc{S-RACP}\xspace}%
	{\textsc{Single-Resource Acquirement Cost Problem}\xspace#1}%
}

\newcommand{\rush}[1][noopt]{\ifthenelse{\equal{#1}{noopt}}{RUSH\xspace}{RUSH$\langle #1 \rangle$\xspace}}
\newcommand{\frush}[1][noopt]{\ifthenelse{\equal{#1}{noopt}}{F-RUSH\xspace}{F-RUSH$\langle #1 
\rangle$\xspace}}

\newcommand{\swag}[1][noopt]{\ifthenelse{\equal{#1}{noopt}}%
	{\textsc{SWAG}\xspace}%
	{\textsc{Scheduling With Augmented Graphs}\xspace}%
}

\newcommand{\grasp}[1][]{\ifthenelse{\equal{#1}{*}}{\textsc{Greedy
Randomized Adaptive Search Procedure}\xspace}{\textsc{GRASP}\xspace}}

\definecolor{color1}{cmyk}{1.0,0,0.6,0} 
\definecolor{color2}{cmyk}{0.8,0.5,0,0} 
\definecolor{color3}{cmyk}{0.25,1,0,0} 

\makeatletter
\newcounter{rfs@totalnumber}
\newcounter{rfs@topnumber}
\newcounter{rfs@dbltopnumber}
\newcounter{rfs@bottomnumber}

\newcommand*{\relaxfloatrestrictions}{%
   \let\rfs@textfraction\textfraction
   \let\rfs@dbltopfraction\dbltopfraction
   \let\rfs@topfraction\topfraction
   \let\rfs@bottomfraction\bottomfraction
   \setcounter{rfs@totalnumber}{\value{totalnumber}}%
   \setcounter{rfs@topnumber}{\value{topnumber}}%
   \setcounter{rfs@dbltopnumber}{\value{dbltopnumber}}%
   \setcounter{rfs@bottomnumber}{\value{bottomnumber}}%

   \renewcommand{\textfraction}{0}%
   \renewcommand{\dbltopfraction}{1.0}%
   \renewcommand{\topfraction}{1.0}%
   \renewcommand{\bottomfraction}{1.0}%
   \setcounter{totalnumber}{100}%
   \setcounter{topnumber}{100}%
   \setcounter{bottomnumber}{100}%
   \setcounter{dbltopnumber}{100}%
}

\newcommand*{\tightenfloatrestrictions}{%
   \let\textfraction\rfs@textfraction
   \let\dbltopfraction\rfs@dbltopfraction
   \let\topfraction\rfs@topfraction
   \let\bottomfraction\rfs@bottomfraction
   \setcounter{totalnumber}{\value{rfs@totalnumber}}%
   \setcounter{topnumber}{\value{rfs@topnumber}}%
   \setcounter{topnumber}{\value{rfs@dbltopnumber}}%
   \setcounter{bottomnumber}{\value{rfs@bottomnumber}}%
}
\makeatother

\usepackage{floatrow}
\newfloatcommand{capbtabbox}{table}[][\FBwidth]

\begin{document}

\title{Engineering Top-Down Weight-Balanced Trees\thanks{This work was supported by the German Research Foundation (DFG) as part of the Research Training Group GRK 2153: Energy Status Data –-- Informatics Methods for its Collection, Analysis and Exploitation.}}

\author{\Large Lukas Barth\thanks{Karlsruhe Institute of Technology, Germany, \{firstname\}.\{lastname\}@kit.edu}
  Dorothea Wagner\footnotemark[2]}

\date {}

\maketitle


\begin{abstract}
  Weight-balanced trees are a popular form of self-balancing binary search trees. Their popularity is due to desirable guarantees, for example regarding the required work to balance annotated trees.

  While usual weight-balanced trees perform their balancing operations in a bottom-up fashion after a modification to the tree is completed, there exists a top-down variant which performs these balancing operations during descend. This variant has so far received only little attention. We provide an in-depth analysis and engineering of these top-down weight-balanced trees, demonstrating their superior performance. We also gaining insights into how the balancing parameters necessary for a weight-balanced tree should be chosen --- with the surprising observation that it is often beneficial to choose parameters which are not feasible in the sense of the correctness proofs for the rebalancing algorithm.
\end{abstract}

\section{Introduction}

Weight-balanced trees (\emph{WBT}s), originally introduced as \emph{binary search trees of bounded balance} or \emph{BB}[$\alpha$]-trees by Nievergelt and Reingold~\cite{Nievergelt1973Binary}, later gained more attention through the seminal work by Knuth~\cite{Knuth1998Art}, which also coined the name \emph{weight-balanced trees} that is better known today. WBTs are balancing binary search trees. As many other flavours of balancing binary search trees, they employ rotations to correct imbalances caused by modifications to the tree. The specialty of weight-balanced trees is that the balancing is done based on the \emph{weight} of subtrees, which is the number of nodes in the respective subtree.

This entails some interesting properties, such as the fact that it can be shown that rotations around \emph{heavy} nodes, \ie, nodes that are roots of subtrees of a large weight, occur only rarely (see Mehlhorn~\cite{Mehlhorn1984Data}). Using this analysis, weight-balanced trees can serve as basis for augmented binary search trees, \ie, trees that carry additional annotations at every node. Usually, e.g.\ in the case of dynamic segment trees, these annotations depend on a node's children, thus the annotation must be repaired if the children are changed. If the effort to repair the annotation at a node after rotation correlates with the weight of the subtree rooted in that node, weight-balanced trees can be used to show amortized bounds on the necessary work. Annotated trees that require this property are often used in computational geometry, examples include Kurt Mehlhorn's Segment Trees (Mehlhorn~\cite[VIII.5.1.3]{Mehlhorn1984Multi}) or Interval Trees (Mehlhorn~\cite[VIII.5.1.1]{Mehlhorn1984Multi}). Also, the weight annotation that every node in a weight-balanced tree carries can be used to efficiently implement order statistic trees (Cormen et al.~\cite[Chapter 15.1]{Cormen1989Introduction}). 

These advantages of weight-balanced trees have led to them receiving ample attention throughout the literature. Adams~\cite{adams_functional_1993} gives a functional implementation of weight-balanced trees and claims they perform as well as red-black trees, however does not provide a practical evaluation. With weight-balanced trees, a set of \emph{balancing parameters} (see Section~\ref{sec:spwbt:wbt}) play a crucial role. While Nievergelt and Reingold introduced the technique and conjectured its correctness, the balancing technique does not work for the whole range of balance parameters they state in their paper. Later, Blum and Mehlhorn~\cite{DBLP:journals/tcs/BlumM80} not only point out this incorrectness, but also give a rigorous proof for a smaller space of the balancing parameters. Hirai and Yamamoto~\cite{hirai2011balancing} use a computer-assisted proof system to discover the whole space of feasible balancing parameters. Cho and Sahni~\cite{Cho2000New} present a variation of weight-balanced trees, which rotates subtrees even if they are not out of balance if the rotation reduces path lengths, thus reducing the expected average node depths within the tree. Roura presents two variations, one that uses logarithmic subtree sizes for balancing~\cite{DBLP:conf/icalp/Roura01} and one that uses the inverse of the Fibonacci function for balancing~\cite{ROURA201348}.

This work focuses on analyzing the advantages of two variations in the weight-balanced trees: first, using a top-down balancing scheme, \ie, repairing the balance constraint while descending the tree for an insertion (resp.\ removal), instead of having a second bottom-up pass over the traversed tree path. Second, the effect that the choice of balancing parameters (especially ``infeasible'' parameters) has. The idea of top-down rebalancing has also been explored for other types of balanced binary search trees, such as red-black trees (Tarjan~\cite{Tarjan1985Efficient}) or weak AVL trees (Haeupler et al.~\cite{Haeupler2015Rank}). Rebalancing weight-balanced trees from the top down has an interesting history: While the original proposal (although incorrect, as Blum and Mehlhorn have shown) by Nievergelt and Reingold was already a top-down algorithm, the supplied proof by Blum and Mehlhorn only works for a bottom-up rebalancing. Later, Lai and Wood~\cite{DBLP:journals/ijfcs/LaiW93} have provided a top-down rebalancing algorithm and shown its correctness. This is the foundation for our contribution. However, the top-down variant of weight-balanced trees has received little attention so far. To our knowledge, no empirical analysis of top-down weight-balanced trees has been done yet.

\paragraph*{Our Contribution}

In this paper, we provide a comprehensive experimental evaluation of top-down as well as bottom-up weight-balanced trees and the possible choices for the balancing parameters, resulting in recommendations when to use which tree variant based on the expected usage pattern. We gain the insight that top-down weight-balanced trees should be preferred over bottom-up weight-balanced trees, and most of the time they can compete with the performance of red-black trees. Moreover, we gain the surprising insight that regarding the choice of balancing parameters, it often is beneficial to chose parameters that violate the theoretical guarantees in favor of a better empirical balance. We also publish thoroughly engineered implementations of all evaluated trees.

\section{Top-Down Weight-Balanced Trees}

In this section, we describe the top-down balancing approach for weight-balanced trees. We start by introducing notation and recapitulating the workings of bottom-up weight-balanced trees in Section~\ref{sec:spwbt:wbt}.

\subsection{Weight-Balanced Trees}%
\label{sec:spwbt:wbt}

We denote a tree $T$ with node set $V$ and edge set $E$ as $T = (V,E)$. Every node $v$ can have a left (resp.\ right) child, which we denote by $L(v)$ (resp. $R(v)$), and say $L(v) = \bot$ (resp. $R(v) = \bot$) if $v$ has no left (resp.\ right) child. Additionally, in a weight-balanced tree, each node has an associated \emph{weight}.  Note that different notions as to what the weight of a node is are found throughout the literature. For us, the weight of $v$, denoted as $|v|$, is the number of nodes in the subtree rooted in $v$ plus one.\footnote{Note that this corresponds to the number of $\bot$ entries in the subtree rooted in $v$.} Thus, a leaf has weight $2$. Also, since in the case $L(v) = \bot$ the left subtree has zero nodes, it results that $|L(v)| = 1$. 

The \emph{balance criterion} for weight-balanced trees limits the relative difference between the weight of the left subtree and the right subtree at every node. The balance criterion and the balancing mechanism use two parameters, $\langle \Delta, \Gamma \rangle$.\footnote{We are using the notation from Hirai and Yamamoto~\cite{hirai2011balancing}.} Balance is achieved at node $v$ if both

\begin{align}
  |L(v)| \cdot \Delta & \geq |R(v)| \text{ and } \label{eq:balance-1} \\
  |R(v)| \cdot \Delta & \geq |L(v)| \text{.} \label{eq:balance-2} 
\end{align}

\noindent Note that the $\Gamma$ parameter is not directly relevant for the balance criterion. If the balance criterion is violated during a modification of the tree, the $\Gamma$ parameter is used to determine the correct balancing procedure. In \cite{DBLP:journals/tcs/BlumM80}, Blum and Mehlhorn show that if $\langle \Delta, \Gamma \rangle$ are chosen in a particular way, and rotations are applied as described in \cite{Nievergelt1973Binary}, this balance criterion is an invariant of the data structure at every node. The proof is technical and tedious, so we do not summarize it here.

\paragraph*{Insertion and Deletion in Bottom-Up Weight-Balanced Trees}

The first pass for insertion and deletion in bottom-up weight-balanced trees is performed as with unbalanced binary search trees. For an insertion, follow the search path for the new node until you walk out of a leaf. This is the position where to insert the new node. For a deletion of $v$, if $v$ is a leaf, just delete it. If it has only one child, replace $v$ by its only child, splicing the node out of the tree. Otherwise, find the largest node in $L(v)$ (resp. the smallest node in $R(v)$), and swap $v$ with that node. Now, $v$ has at most one child and we can proceed as above. The insertion procedure is also shown in Algorithm~\ref{alg:insertion}.

During the above, we do not pay attention to any balance criterion. Thus, after insertion and deletion, the balance might be violated at several nodes on the path from the tree's root to the position of insertion or deletion. In bottom-up weight-balanced trees, we repair the tree by traversing that path back up, repairing imbalances using single rotations and double rotations as necessary.

\paragraph*{Rebalancing Operation}

\begin{figure*}
  \centering
  \includegraphics{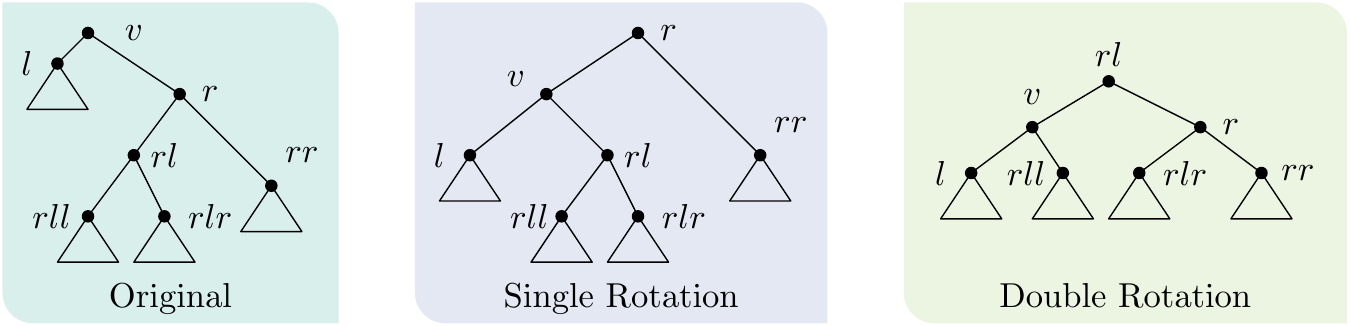}
  \caption{The result of a single left rotation around $v$ and a double rotation, first right around $r$, then left around $v$. Note that the node names differ from the notation in text to provide consistent labels before and after rotation. Also, the notation is reversed from the function notation to yield more natural node names. For example, $rll$ corresponds to $L(L(R(v)))$. Triangles indicate (possibly empty) subtrees that have been omitted.}
  \label{fig:rotations}
\end{figure*}

Whenever the balance criterion at a node $v$ is violated, a single or double rotation as depicted in Figure~\ref{fig:rotations} is performed to reestablish balance. Since the process is symmetric for the right and left subtrees, we only discuss the case that the right subtree has become too heavy (because of an insertion into $R(v)$ or a deletion from $L(v)$).

Given a $\langle \Delta, \Gamma \rangle$ pair as defined above, the first decision at $v$ is whether to perform a rotation at all. A rotation is performed if (in the case of a possible right-overhang) $|L(v)| \cdot \Delta < |R(v)|$. A left rotation around $v$ will certainly reduce the weight of $v$'s right subtree, essentially removing $R(v)$ and the subtree rooted in $R(R(v))$ from below $v$. However if $L(R(v))$ is too heavy, after the rotation, the balance at the old $R(v)$ could be violated. Thus, if $|L(R(v))| > |R(R(v))| \cdot \Gamma$, we perform a double rotation as shown in Figure~\ref{fig:rotations}.\footnote{Note that the node names in the figure differ from the node names in text to allow for consistent names before and after rotation.} This procedure has been shown to always reestablish balance at all involved nodes by Blum and Mehlhorn if $\langle \Delta, \Gamma \rangle$ is chosen appropriately.

\paragraph*{Balancing Parameter Space}

When talking about the balancing parameters $\langle \Delta, \Gamma \rangle$, we often call them \emph{feasible} or \emph{infeasible}. A parameter set $\langle \Delta, \Gamma \rangle$ is \emph{feasible} (for bottom-up rebalancing resp. top-down rebalancing) if the respective balancing algorithm has been shown to be correct for $\langle \Delta, \Gamma \rangle$, \ie, if it is guaranteed that all nodes satisfy (\ref{eq:balance-1}) and (\ref{eq:balance-2}) after rebalancing. Otherwise, the parameter set is called \emph{infeasible}. Note that an infeasible parameter set still yields a valid binary search tree.

Regarding the feasible values for $\langle \Delta, \Gamma \rangle$, the first thing to note is that the two correctness proofs from Blum and Mehlhorn as well as Lai and Wood~\cite{DBLP:journals/ijfcs/LaiW93} use a different notation than $\langle \Delta, \Gamma \rangle$. In these proofs, the balancing factor is $\alpha$, and the balancing criterion is

$$\alpha \leq \frac{|L(v)|}{|L(v)| + |R(v)|} \leq (1 - \alpha)$$

\noindent Looking at only one side of both types of balance constraints (the other side is symmetric), from $|L(v)| / (|L(v)| + |R(v)|) \geq \alpha$ and $\Delta |L(v)| \geq |R(v)|$, we get that $\Delta = (1 - \alpha) / \alpha$. In fact, using the upper bound on $\alpha$ given by Blum and Mehlhorn, $\alpha \leq 1 - \sqrt{2} / 2$, this leads to $\Delta \geq 1 + \sqrt{2}$. Note that the larger the value for $\alpha$ (and the smaller the value for $\Delta$), the better we expect the tree to be balanced, \ie, we expect the smallest average node depths for these values. For their correctness proofs, both Blum and Mehlhorn as well as Lai and Wood fix the second balance parameter (the parameter deciding whether to use single or double rotation, call it $\gamma$) to $\gamma = 1/(2 - \alpha)$. Again, taking the two different forms of constraints for a double rotation, namely $|L(v)| > \Gamma |R(v)|$ and $|L(v)| / (|L(v)| + |R(v)|) > \gamma$, it follows that $\Gamma = \gamma / (1 - \gamma)$ and therefore $\Gamma = 1 / (1 - \alpha)$. With this, for $\alpha = 1 - \sqrt{2}/ 2$, it follows that $\Gamma = \sqrt{2}$, and with that the most common (and maximally balanced) choice for $\langle \Delta, \Gamma \rangle = \langle 1 + \sqrt{2}, \sqrt{2} \rangle$. 

However, Hirai and Yamamoto~\cite{hirai2011balancing} have shown that in the bottom-up balancing case, the feasible space for $\langle \Delta, \Gamma \rangle$ is in fact a nonempty polytope, \ie, the linear dependency between $\Delta$ and $\Gamma$ (resp. $\alpha$ and $\gamma$) is not necessary. The only integral choice for $\langle \Delta, \Gamma \rangle$ within the polytope is $\langle 3, 2 \rangle$. Integral values for $\langle \Delta, \Gamma \rangle$ are interesting since (as Roura~\cite{DBLP:conf/icalp/Roura01} shows), using floating point arithmetic, or even worse, computing $\sqrt{2}$ during balancing, is a major factor slowing down weight-balanced trees. Note that with the relationship between $\Delta$ and $\Gamma$ (resp. $\alpha$ and $\gamma$) established by Blum and Mehlhorn, the $\Gamma$ value for $\Delta = 3 $ would have been $\Gamma = 4/3$.

The correctness proof for top-down balancing from Lai and Wood holds only for $\alpha \leq 1/4$, meaning that we expect the best balanced top-down weight-balanced trees for $\alpha = 1/4$, which translates to $\langle \Delta, \Gamma \rangle = \langle 3, 4/3 \rangle$. Note that even though this means that $\Delta = 3$ is feasible for top-down balancing, the aforementioned $\langle 3, 2 \rangle$ possibly is not a feasible choice in the top-down case, since it is unclear how the feasible polytope looks like.

\subsection{From Bottom-Up to Top-Down}

Weight-balanced trees as described above perform two full traversals of the path from the tree's root to a leaf (resp. to-be-deleted node) for each insertion and deletion: One traversal down to perform the deletion or insertion, and one traversal up to check for and repair the balance. However, whenever we know that we will definitely delete a node (e.g., because we know that the value to be deleted is in the set represented by the tree), or that we will definitely insert a node (e.g., because we allow multiple nodes with the same value to be inserted), it is possible to perform necessary repair operations on the first traversal towards the leaves. 

Algorithm~\ref{alg:insertion} shows pseudocode for such an insertion. Note that while \texttt{Insert} descends the tree towards the insertion position for $n$, \texttt{RepairDuringInsertion} is called at every node, performing rotations as if $n$ was already inserted into the appropriate subtree, but without that subtree being rebalanced before. The pseudocode omits some technical details, such as correctly adjusting the weights of the nodes that become ancestors of $v$ because of a rotation, and correctly descending in case of a rotation. Consider as an example for a more complicated procedure the case that $n > v$, $n > R(v)$, that the insertion causes $|R(v)| > |L(v)| \cdot \Delta$ and that $|L(R(v))| > |R(R(v))| \cdot \Gamma$. Then, \texttt{RepairDuringInsertion} calls a double rotation (see Figure~\ref{fig:rotations}), after which $n$ should of course still be inserted below the old $R(R(v))$ ($rr$ in Figure~\ref{fig:rotations}). However, that node is not a descendant of $v$ anymore.

Note that this approach does not lead to the same trees as the bottom-up approach. In the bottom-up approach, during rebalancing at node $v$, balance is already established at $L(v)$, $R(v)$ and all nodes below. In the top-down approach, this balance can be violated by up to one node. For the top-down procedure, Lai and Wood show that even though the lower nodes cannot yet be assumed to be balanced, the above procedure balances all involved nodes, if $\langle \Delta, \Gamma \rangle$ is chosen appropriately.

The approach outlined here assumes that every insertion and removal always changes the tree. This is not necessarily the case, as a removal of a value that is not in the tree will fail, and so will insertion if the tree is used to implement a set (instead of a multiset) and the value is already in the tree. The case that the tree is not modified can naively be accommodated by having a second pass over the modified path in that case. Obviously, with this naive solution, top-down rebalancing is only a useful approach if the number of modifying insertions and removals is way higher than the number of non-modifying ones. However, careful analysis by Lai and Wood~\cite{DBLP:journals/ijfcs/LaiW93} shows that for a correct choice of rebalancing parameters, top-down rebalancing keeps the balancing criterion intact even if the algorithm aborts the operation during descend, either because a key to be deleted is not in the tree or because a key to be inserted is already in the tree.\footnote{Lai and Wood call this a \emph{redundant} operation.}

\begin{algorithm}[t]
  \SetKwFunction{FnRepair}{RepairDuringInsertion}
  \SetKwFunction{FnInsert}{Insert}
  \SetKwProg{Fn}{Function}{:}{}
  
  \Fn{\FnRepair{$n$, $v$}} {
  \eIf{$n \leq v$}{
    \If(\tcp*[f]{$+1$ b/c we insert into left subtree}){$|L(v)| + 1 > |R(v)| \cdot \Delta$}{
      \eIf{$(n \leq L(v)$ \textbf{and} $|R(L(v))| > (|L(L(v))| + 1) \cdot \Gamma)$ \textbf{or}\\ \hspace{0.41cm}$(n > L(v)$ \textbf{and} $|R(L(v))| + 1 > |L(L(v))| \cdot \Gamma))$}{
        doubleRotation(v)\;
      }{
        singleRotation(v)\;
      }
    }
  }{
    \tcc{Omitted, symmetric to the case $n \leq v$}
  }
  }
  \vspace{0.3cm}
  \Fn{\FnInsert{$n$}} {
    $v \leftarrow root$\;
    \While{true} {
      $|v| \leftarrow |v| + 1$\;
      \FnRepair($n$, $v$)\;
      \eIf{$n \leq v$} {
        \eIf{$L(v) = \bot$} {
          $L(v) \leftarrow n$\;
          \KwRet;
        }{
          $v \leftarrow L(v)$\;
        }
      }{
        \tcc{Omitted, symmetric to the case $n \leq v$}
      }        

    }
  }
  \caption{Top-down insertion of a node $n$.}
  \label{alg:insertion}
\end{algorithm}%

\section{Evaluation}

We now provide an in-depth experimental evaluation of the various flavours of weight-balanced trees. This evaluation encompasses multiple parts: First, we measure the time that operations such as inserting into and removing from the trees take in Section~\ref{sec:timing}. Since the time necessary to search for a vertex in a tree is only dependent on the depth of the respective node, and measuring the average node depth is less noisy than measuring the time a search takes, we use this measure instead of measuring search timings in Section~\ref{sec:balance}. Also in that section we look at how much the balancing criterion is violated when one chooses balancing parameters outside the feasible space. All these analyses are done for different kinds of test data, resembling a broad spectrum of use cases. To study the various rebalancing schemes in even more realistic scenarios, we use sequences of tree operations captured during the execution of an optimization algorithm utilizing a balancing search tree in Section~\ref{sec:real-life}. Finally, we take a look at the total number and weight of rotated nodes in Section~\ref{sec:rotated-weight}.

We implemented all trees in C++, our implementation including all the benchmarking code can be found at:

\begin{center}
  \url{https://github.com/tinloaf/ygg/releases/tag/version_used_for_alenex20}
\end{center}

\noindent Additionally, we publish all raw results obtained from our experiments as a separate data publication~\cite{data-publication}. See Section~\ref{sec:appendix-data} in the appendix for further details.

All measurements are taken on a machine with 192 GBs of DDR4 memory and two eight-core Intel\textsuperscript{\textregistered} Xeon\textsuperscript{\textregistered} Gold 6144 CPUs, which have 32 KB of L1 data cache per core, 1 MB of L2 cache per core and a total of 25 MB of L3 cache per CPU. However, we did not run multiple benchmarks concurrently. The size of our trees' nodes is 40 bytes. In the following experiments, the largest tested trees usually have size $\approx 4 \times 10^6$, which leads to a memory footprint of around $150 \>\textrm{MB}$, well above L3 cache sizes. We therefore expect to see the effects of caching for the larger tested trees, and little to no caching effects for trees of at most $6 \times 10^5 \approx 25 \>\textrm{MB} / 40 \>\textrm{B}$ nodes.


In the following evaluation, we compare the following balanced binary search trees: First, a (bottom-up) red-black tree as baseline, denoted \emph{red-black}. Second, the basic version of a weight-balanced tree, with bottom-up balancing, denoted \emph{bottom-up}. Third, the top-down weight-balanced tree, denoted \emph{top-down}. For the top-down weight-balanced trees, we evaluate different choices for the balancing parameters $\langle \Delta, \Gamma \rangle$: First, the choices listed and explained in Section~\ref{sec:spwbt:wbt}: $\langle 1 + \sqrt{2}, \sqrt{2} \rangle$ (the original parameter set given by Blum and Mehlhorn~\cite{DBLP:journals/tcs/BlumM80}), $\langle 3, 2 \rangle$ (the integral parameters suggested by Hirai and Yamamoto~\cite{hirai2011balancing}) and $\langle 3, 4/3 \rangle$ (the parameters from the top-down correctness proofs by Lai and Wood~\cite{DBLP:journals/ijfcs/LaiW93}). Note that even though the first two are not feasible in the sense of the top-down correctness proof by Lai and Wood, we still use them for the top-down balancing technique. Similarly, we try the additional choice of $\langle 2, 3/2 \rangle$. Even though $\Delta = 2$ is not a feasible choice for top-down \emph{or} bottom-up balancing, we want to evaluate how this smaller $\Delta$ value (which we expect to lead to a better balance) performs in practice. For an even more extreme example, we also evaluate $\langle 3/2, 5/4 \rangle$.

\subsection{Timing Operations}
\label{sec:timing}

We first benchmark the two basic operations insertion and deletion. Our aim is to measure the time these operations take on trees of various sizes for different distributions of nodes' keys. Specifically, for each benchmark we first create a random tree of a certain base size (the \emph{base tree}), and then remove five percent of the nodes resp.\ insert five percent new nodes. For all benchmarks, we employ four different methods to generate nodes' keys: First in the \texttt{uniform} case, we generate keys uniformly at random. Second, we assume that the search tree may be used to index data that pertains to physical or social sciences. In this case, Zipf's Law (see~\cite{powers1998applications}) states that this data, \eg{} text corpora, often follow a Zipf distribution. We accommodate this fact with the \texttt{zipf} case, in which nodes' keys are picked using a Zipf distribution. Third, it seems prudent to study cases with a heavy concentration of the keys in one or two areas of the key space. For this, we use the \texttt{skewed} distribution suggested by M{\"a}kinen~\cite{Mäkinen1987} in his analysis of top-down splay trees. In this distribution, every third value is drawn from a uniform distribution over the whole key space, and the other two thirds are drawn from two uniform distributions each spanning only $10 \%$ of the key space. Finally, an obvious benchmark case for balancing search trees is partially pre-sorted data. In the \texttt{pre-sorted} case, we first take a sequence of sorted numbers, and then randomly permute half of them. For the deletion benchmark, the node to be deleted is picked uniformly at random in each case.

Note that we do not discuss each plot individually in this section, but only those from which interesting insights can be drawn. The plots that are not mentioned in the text can be found in Appendix~\ref{app:omitted-benchmark}.

To account for randomness effects and measurement noise, we run each experiment for each base tree size on $10$ different base trees, and in turn repeat the experiment itself on each base tree until the experiment ran for at least one second on each base tree. 

\begin{figure*}
  \centering
  \begin{subfigure}[t]{0.39\textwidth}
    \includegraphics[width=0.95\textwidth]{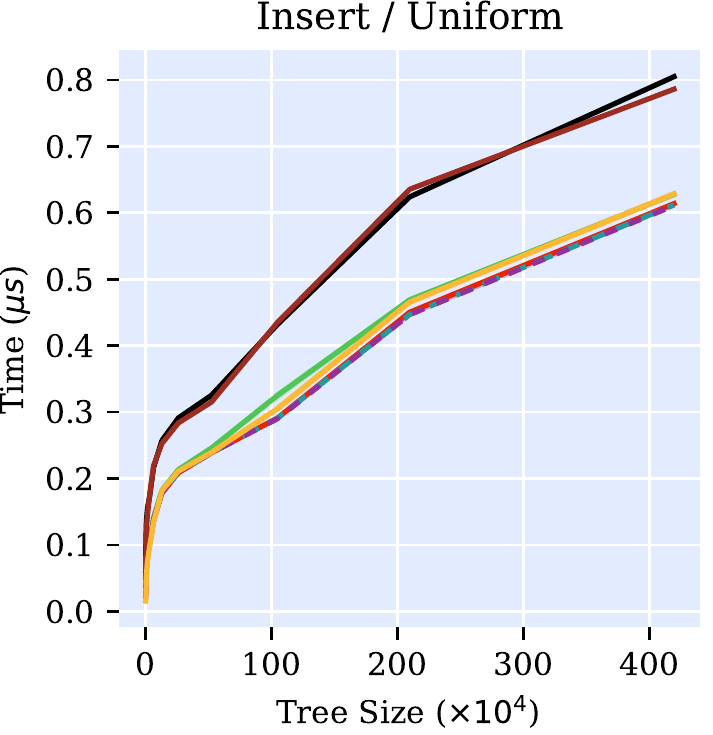}
    \caption{Nodes' keys chosen from a uniform distribution.}%
    \label{plt:insert-time-uniform}
  \end{subfigure}
  \quad
  \begin{subfigure}[t]{0.57\textwidth}
    \includegraphics[width=0.95\textwidth]{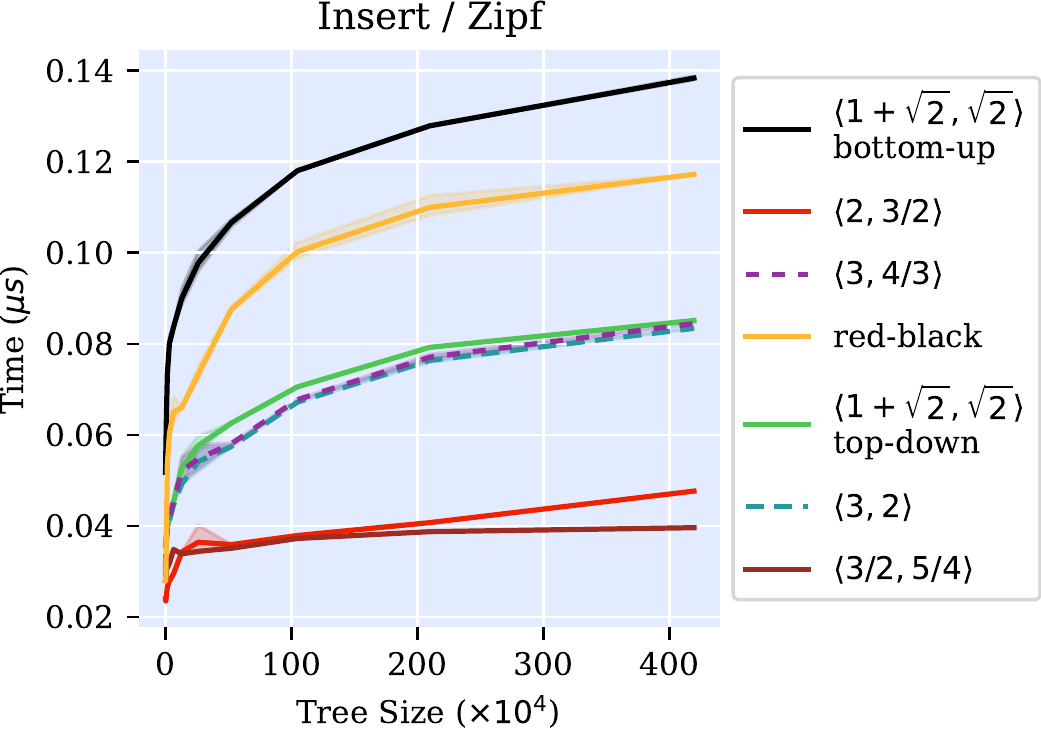}
    \caption{Nodes' keys chosen from a Zipf distribution.}%
    \label{plt:insert-time-zipf}
  \end{subfigure}
    \caption{Times to insert $5\%$ new nodes into trees of various sizes. The $x$ axis specifies the size of the base tree. The $y$ axis reports the time needed for a single insertion in microseconds. Shaded areas indicate standard deviation.}
\end{figure*}


Figure~\ref{plt:insert-time-uniform} shows the time (averaged over all the iterations explained above) it takes to insert $5\%$ new nodes into the seven different trees of various base sizes, for the \texttt{uniform} case. We first see that the bottom-up variant (with $\langle 1 + \sqrt{2}, \sqrt{2}\rangle$) is about $30 \%$ (resp.\ $0.2 \mu{}s$) slower than the corresponding top-down variant. We then see that the red-black tree and the $\langle 2, 3/2 \rangle$, $\langle 3, 4/3 \rangle$ and the top-down $\langle 1 + \sqrt{2}, \sqrt{2}\rangle$ variants all show almost the same performance. Interestingly, the variant with the tightest balancing parameter, $\langle 3/2, 5/4 \rangle$ performs as bad as the bottom-up variant. Note that in this (and all following) plots, shaded areas indicate standard deviation. Where no shaded area is visible, the standard deviation is too small to be visible.

When performing the same experiment with node keys chosen from a Zipf distribution (shown in Figure~\ref{plt:insert-time-zipf}), results look very different: Here, the two strongly balanced variants ($\langle 2, 3/2 \rangle$ and $\langle 3/2, 5/4\rangle$) outclass all other variants. Also, all but the top-down variant outclass the red-black tree, with a factor of $3$ between the best balanced weight-balanced tree and the red-black tree. Results for the \texttt{skewed} distribution (shown in Figure~\ref{plt:insert-time-skewed} in Appendix~\ref{app:omitted-benchmark}) are less pronounced, but similar. For the \texttt{pre-sorted} case, the results are very similar to the \texttt{uniform} case and can be found in Figure~\ref{plt:insert-time-presorted} in Appendix~\ref{app:omitted-benchmark}.

Since we implemented all mentioned trees ourselves, the question of how efficient our implementations are as a whole comes to mind. For the insertion benchmark, we added the C++ STL's \texttt{std::multiset}\footnote{STL bundled with GCC 8.1, which implements std::multiset as a red-black tree.} and Boost's \texttt{intrusive::multiset} to the comparison. The plot can be found in Appendix~\ref{app:omitted-benchmark}, Figure~\ref{plt:comparison-stdset}. As can be seen, all our trees perform slightly better than \texttt{std::multiset}, but slightly worse than \texttt{boost::intrusive::multiset}. We may therefore assume that our implementations are properly optimized.

\begin{figure*}
  \centering
  \begin{subfigure}[t]{0.40\textwidth}
    \includegraphics[width=0.95\textwidth]{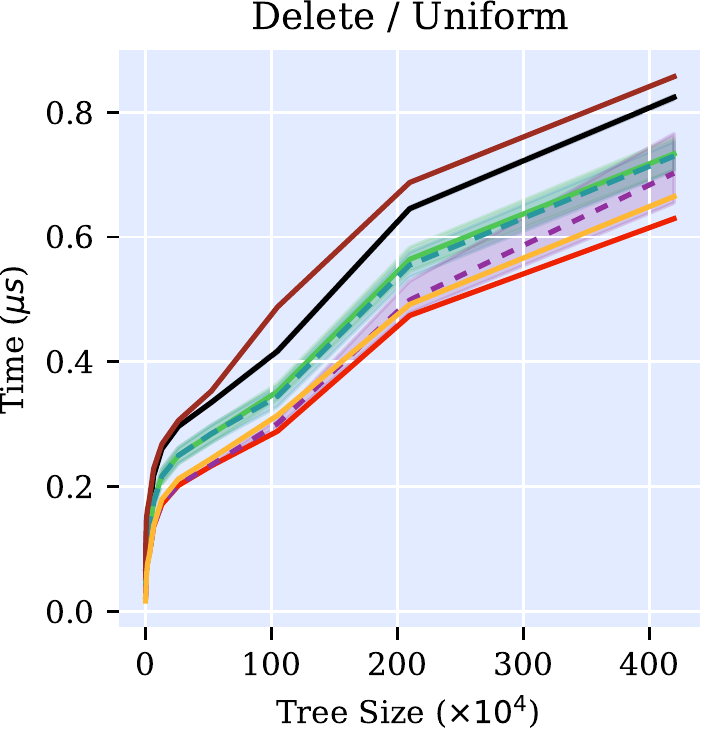}
    \caption{Node keys generated from a uniform distribution.}%
    \label{plt:erase-uniform}
  \end{subfigure}
  \quad
  \begin{subfigure}[t]{0.56\textwidth}
  \includegraphics[width=0.95\textwidth]{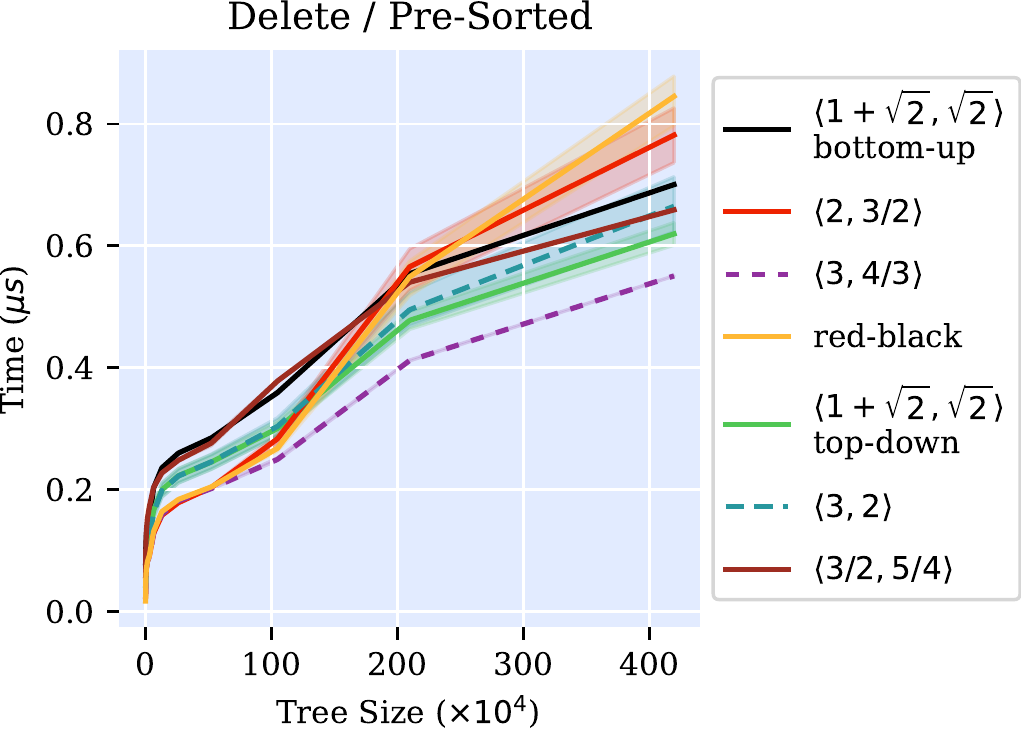}
  \caption{Node keys generated in the \texttt{pre-sorted} fashion.}%
  \label{plt:erase-presorted}

  \end{subfigure}
  
  \caption{Times to delete $5\%$ nodes from trees of various sizes. The $x$ axis specifies the size of the base tree. The $y$ axis reports the time needed for a single deletion in microseconds. Shaded areas indicate standard deviation.}
  
\end{figure*}
\begin{table}
  \begin{tabular}{l@{\hskip 2em}c@{\hskip 2em}c}
    \toprule
    & Deletion & Insertion \\
    \midrule
    \texttt{uniform} & $\langle 2, 3/2 \rangle \checkmark $ & $\langle 3, 4/3 \rangle / \langle 2, 3/2 \rangle \checkmark$ \\
    \texttt{skewed} & $\langle 3, 4/3 \rangle \times$  & $\langle 3, 4/3 \rangle / \langle 2, 3/2 \rangle \times$ \\
    \texttt{zipf} & $\langle 3, 2 \rangle \times $& $\langle 3/2, 5/4\rangle \checkmark$  \\
    \texttt{pre-sorted} & $\langle 3, 4/3 \rangle \checkmark $ & $\langle 3, 4/3 \rangle / \langle 2, 3/2 \rangle \checkmark$ \\
    \bottomrule
  \end{tabular}
  \caption{Summary of the benchmark findings, specifying which weight-balanced tree variant was the best for each of our benchmark cases. Where two variants could virtually not be distinguished, we specify both. A checkmark signifies that in this case, the best weight-balanced tree outperformed the red-black tree, a cross the opposite.}
  \label{tab:bench-summary}
\end{table}

\begin{figure*}
  \centering
  \begin{subfigure}[t]{0.40\textwidth}
    \includegraphics[width=0.95\textwidth]{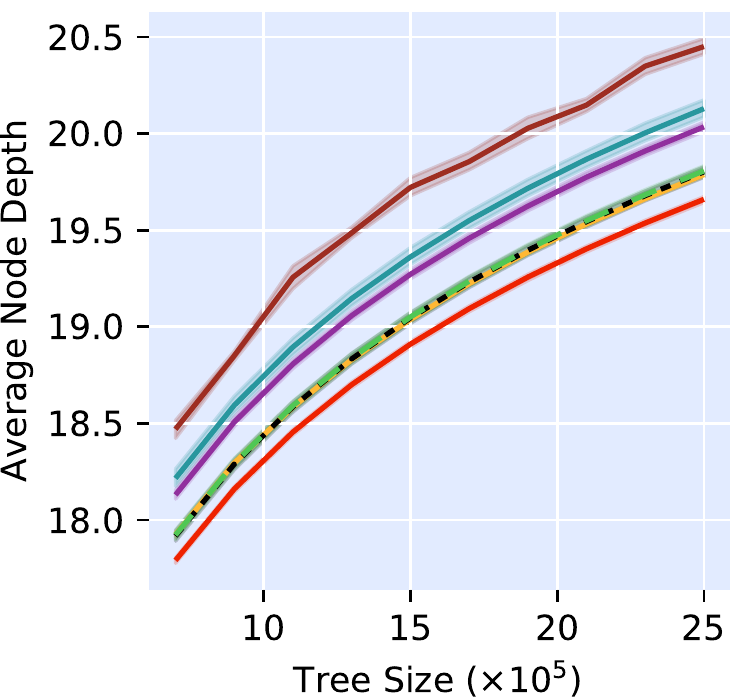}
    \caption{Node keys generated from a uniform distribution.}%
    \label{plt:average-depth-uniform}
  \end{subfigure}
  \quad
  \begin{subfigure}[t]{0.56\textwidth}
  \includegraphics[width=0.95\textwidth]{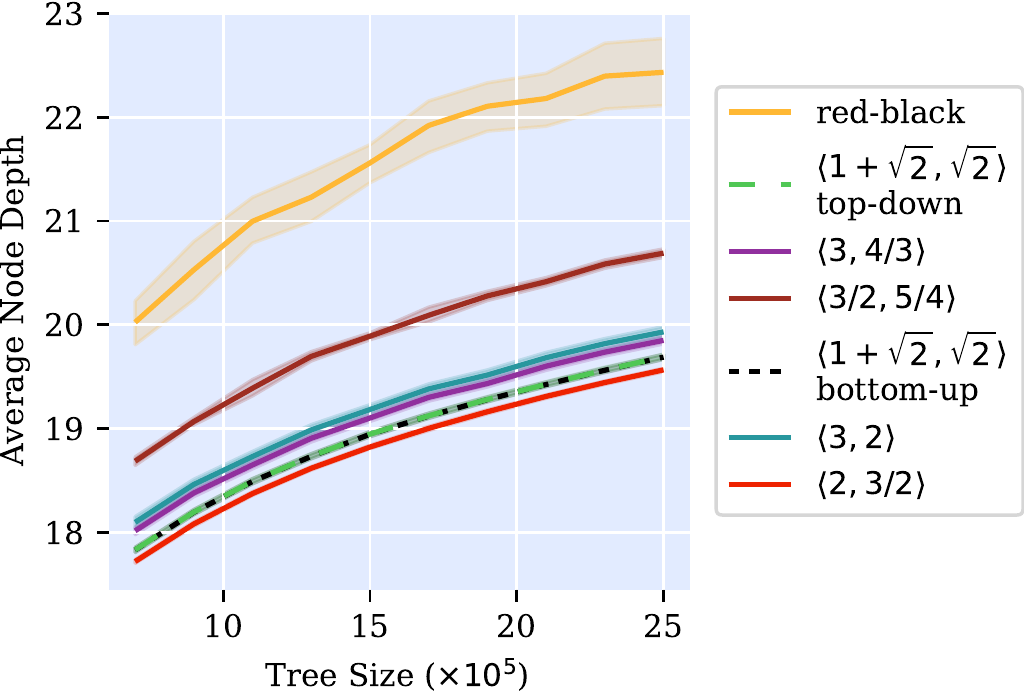}
  \caption{Node keys generated from a Zipf distribution.}%
  \label{plt:average-depth-zipf}
  \end{subfigure}
    \caption{Average node depth for various trees. The $x$ axis specifies the size of the tree, the $y$ axis the average node depth. All nodes in every tree were randomly generated, removed once, had their key changed, and were reinserted. The solid lines indicate average values, the shaded areas the standard deviation.}
  \label{plt:average-depth}
\end{figure*}

\begin{figure}[t]
  \includegraphics[width=0.95\textwidth]{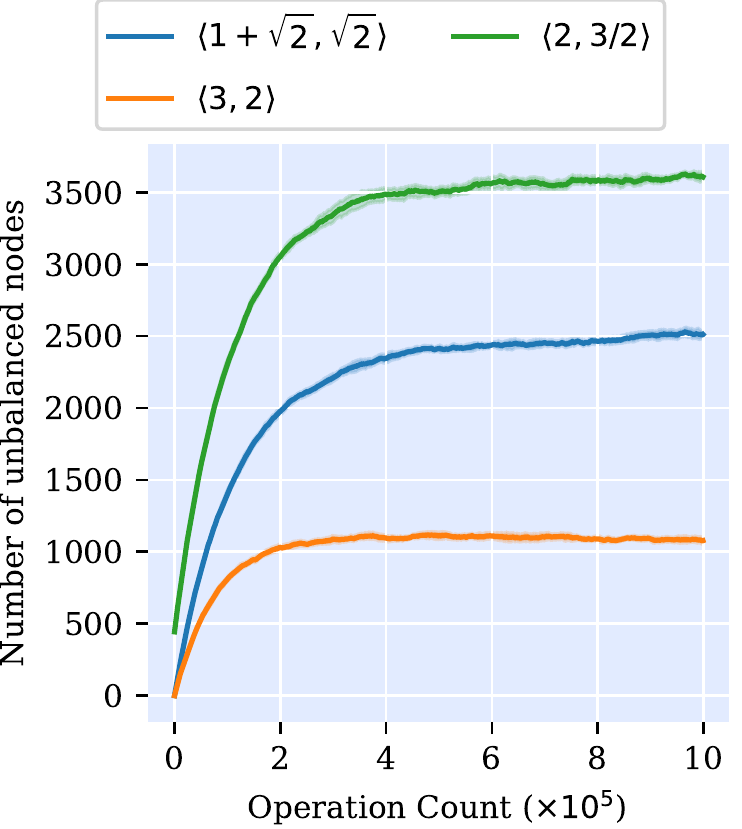}
  \caption{Number of unbalanced nodes on the $y$ axis versus number of remove / insert operations on random trees of size $10^6$ on the $x$ axis. The solid line reports the mean value, the shaded area indicates the standard deviation.}
  \label{plt:violations-over-time}
\end{figure}

\begin{figure*}[t]
  \centering
  \includegraphics[width=0.85\textwidth]{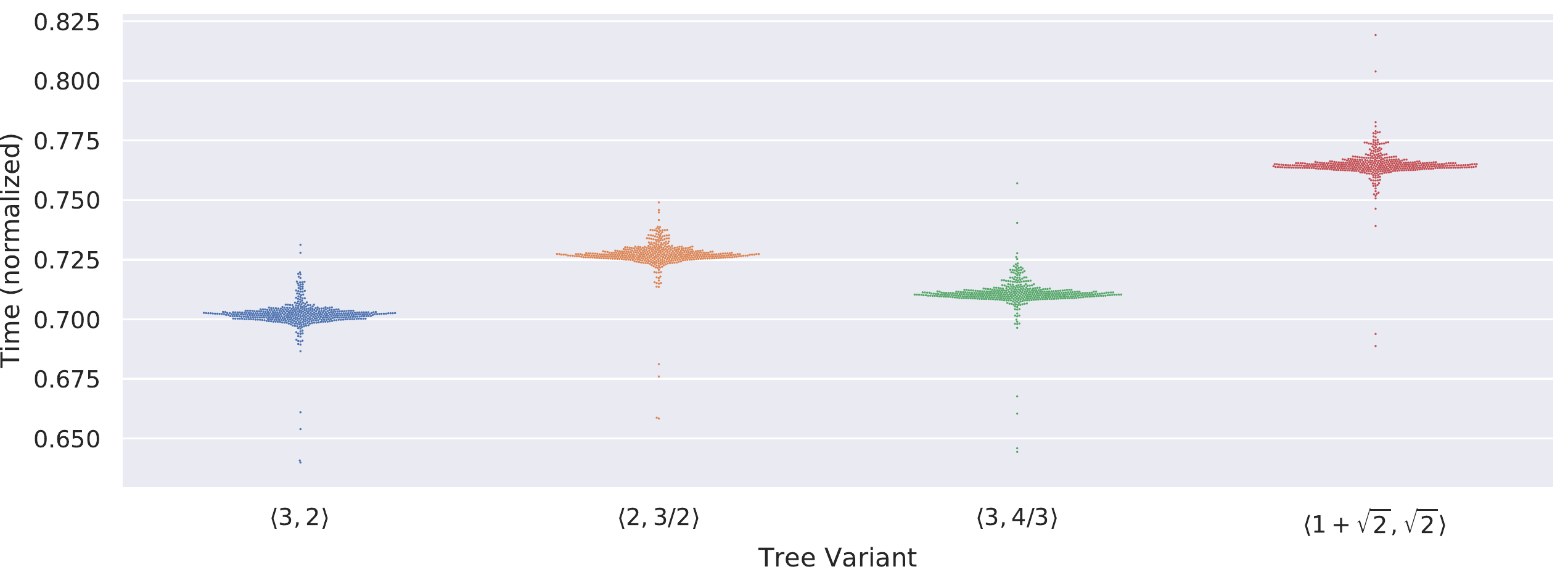}
  \caption{Times elapsed during the execution of each captured sequence, normalized to the time it took the bottom-up weight balanced tree with $\langle 1 + \sqrt{2}, \sqrt{2} \rangle$. Every dot is one sequence.}
  \label{plt:sequence-timing}
\end{figure*}

Next, we look at the deletion operation. Figure~\ref{plt:erase-uniform} shows the \texttt{uniform} case. We see that the $\langle 2, 3/2 \rangle$ variant has a slight advantage over the red-black tree and all other variants. For deletion, the \texttt{pre-sorted} case (shown in Figure~\ref{plt:erase-presorted}) is especially interesting: Here, all weight-balanced trees, but especially the $\langle 3, 2\rangle$ variant, clearly outperform the red-black tree. On the other side of the spectrum, for the \texttt{skewed} and \texttt{zipf} cases (shown in Appendix~\ref{app:omitted-benchmark}, Figure~\ref{plt:erase-zipf-skewed}), the red-black tree has a slight advantage over the weight-balanced trees.

Our benchmark findings are summarized in Table~\ref{tab:bench-summary}. From the results, we can deduce that one should always use the top-down variant, and should never use $\langle 1 + \sqrt{2}, \sqrt{2} \rangle$ as balancing parameter. Whether $\Delta = 2$ or $\Delta = 3$ is the wiser choice depends on the expected usage pattern. We can also see that the race between red-black trees and weight-balanced trees is a toss-up: While weight-balanced trees seem to be ahead in the \texttt{uniform} and \texttt{pre-sorted} cases, red-black trees exhibit better performance in the \texttt{skewed} and \texttt{zipf} cases.


\subsection{Tree Balance}
\label{sec:balance}
Aside from insertion and deletion times, an interesting metric is the average depth of a node. The average depth determines the expected length of the search path for that node, which not only influences insertion and removal speeds, but even more strongly the search performance. In fact, we do not benchmark runtimes for searches within the trees, since the average depth of the nodes should be the only influencing parameter, with everything else being measurement noise. To analyze the average node depth, we again create random trees of various sizes. Since just creating a tree does not involve the remove operation, and we also want to evaluate the effects of this operation, we iterate over all nodes after creating the tree, and first remove each node from the tree, change its key, and then reinsert it. After this, we compute the average depth of all nodes. Figure~\ref{plt:average-depth-uniform} shows the results for keys being drawn from a uniform distribution.

We see that red-black trees and weight-balanced trees using $\langle 1 + \sqrt{2}, \sqrt{2} \rangle$ as balance parameters are virtually equally well balanced. The weight-balanced tree using $\langle 2, 2/3 \rangle$ has a slight advantage over them --- as we expected, since $\Delta = 2$ enforces a stricter balance than $\Delta = 1 + \sqrt{2}$. However, the $\langle 3/2, 5/4 \rangle$ variant is the worst in terms of balance, even though it is using the smallest $\Delta$. This suggests that choosing parameters that are too far outside of the space of feasible choices for $\langle \Delta, \Gamma \rangle$, the balancing criterion is violated too badly for the smaller $\Delta$ to make up for it.

Using a Zipf distribution instead of a uniform distribution for the nodes' keys (shown in Figure~\ref{plt:average-depth-zipf}) reveals that while the various weight-balanced trees are almost unaffected by the heavily skewed distribution, the red-black tree handles it a lot worse, with more than $10 \%$ difference between the best weight-balanced tree and the red-black tree. Interestingly, the \texttt{skewed} case, shown in Figure~\ref{plt:average-depth-skewed} in Appendix~\ref{app:omitted-benchmark}, shows results very similar to the \texttt{uniform} case.


The fact that in Figure~\ref{plt:average-depth}, the values for the top-down weight-balanced tree with $\langle 1 + \sqrt{2}, \sqrt{2} \rangle$ do not differ much from the bottom-up weight-balanced tree with the same balancing parameters (which are infeasible for a top-down balancing approach), and that the (infeasible) parameter pair $\langle 2, 3/2 \rangle$ outperforms all other trees, hint at the fact that even infeasible balancing parameters for a top-down balancing approach may produce little to no balance violations in practice. We examine this claim by continually counting the number of nodes at which balance is violated while repeatedly removing and re-inserting (with a changed value) random nodes from resp.\ into a random tree. Figure~\ref{plt:violations-over-time} shows the results for a tree of size $10^6$. We repeat the experiment with $10$ different seeds, the line indicates the mean, the shaded areas indicate the standard deviation. We see that for all three\footnote{We excluded $\langle 3/2, 5/4 \rangle$ here, since it breaks the balancing so badly that it distorts the plot.} evaluated variants, the number of nodes at which balance is violated stabilizes after approximately $4 \times 10^5$ removals and insertions.\footnote{One removal and one insertion count as one operation.} We also see that even for the worst of the parameter choices, $\langle 2, 3/2 \rangle$, only about $0.35\%$ of all nodes are unbalanced after $10^6$ operations. This behavior can be explained by the fact that unbalanced nodes will likely be rebalanced by the next operation that passes over them. Thus, we also expect the unbalanced nodes to have large depths, since nodes close to the root are passed over very frequently.

\subsection{Real-Life Sequences}
\label{sec:real-life}
After the experiments on randomly generated data, we finally take a look at tree operations generated from an algorithm that heavily relies on balancing binary trees. To this end, we instrumented the SWAG algorithm by Barth and Wagner~\cite{barth2019shaving}. This is a scheduling algorithm that in its innermost loop uses a dynamic segment tree, which is built on top of a balancing binary search tree. Note that the algorithm only deletes from and inserts into the tree and never performs any searches.\footnote{While this might seem useless, the information needed by the scheduling algorithm is computed in an annotation at the root of the tree.} We collected a total of $514$ sequences of tree operations. To benchmark our weight-balanced trees, we replay each sequence ten times for every weight-balanced tree variant. Figure~\ref{plt:sequence-timing} shows the results, where every dot is the time (averaged over the ten iterations) it took the tree indicated by the $x$ axis to execute one sequence, normalized to the time it took the bottom-up $\langle 1 + \sqrt{2}, \sqrt{2}\rangle$ variant to execute the same sequence. We see that in this specific use case, the $\langle 3, 2\rangle$ variant suggested by Hirai and Yamamoto (for the bottom-up variant) performs the best, being about $30 \%$ faster than the bottom-up variant --- again, we see the best results for a parameter choice that is infeasible for top-down rebalancing. The top-down feasible variant $\langle 3, 4/3 \rangle$ performs slightly worse. But even the worst variant, $\langle 1 + \sqrt{2}, \sqrt{2} \rangle$ still is more than $23 \%$ faster than the bottom-up variant. Note that we have excluded the $\langle 3/2, 5/4 \rangle$ variant here. It has its mean at approximately $1.15$ and would distort the plot.

\subsection{Rotated Node Weight}
\label{sec:rotated-weight}

\begin{figure*}
  \centering
  \begin{subfigure}[t]{0.39\textwidth}
   \includegraphics[width=0.95\textwidth]{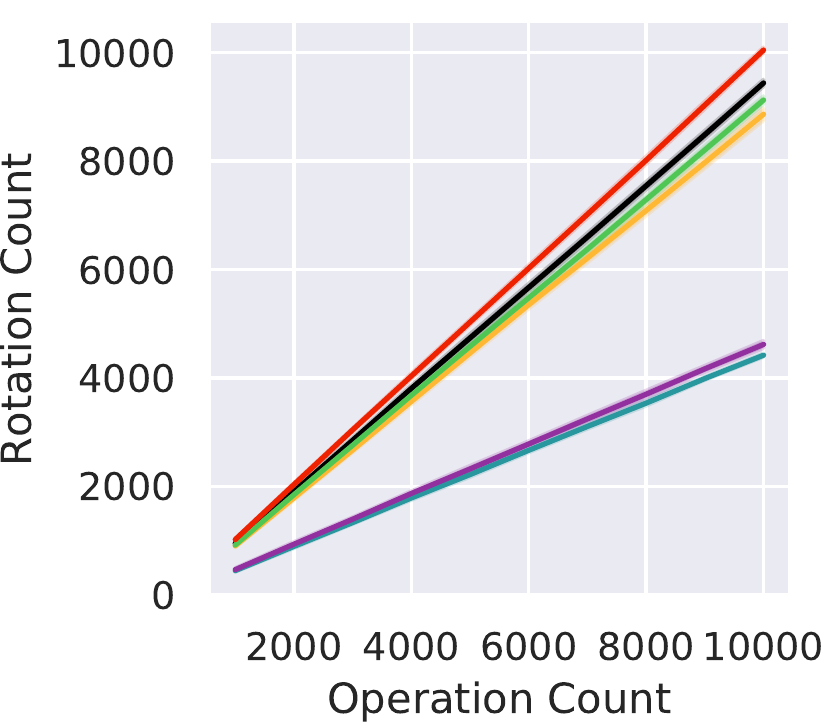}
   \caption{Total number of of performed rotations (on the $y$ axis) after a number of operations (on the $x$ axis).}
    \label{plt:rotation-count}
  \end{subfigure}
  \quad
  \begin{subfigure}[t]{0.57\textwidth}
  \includegraphics[width=0.95\textwidth]{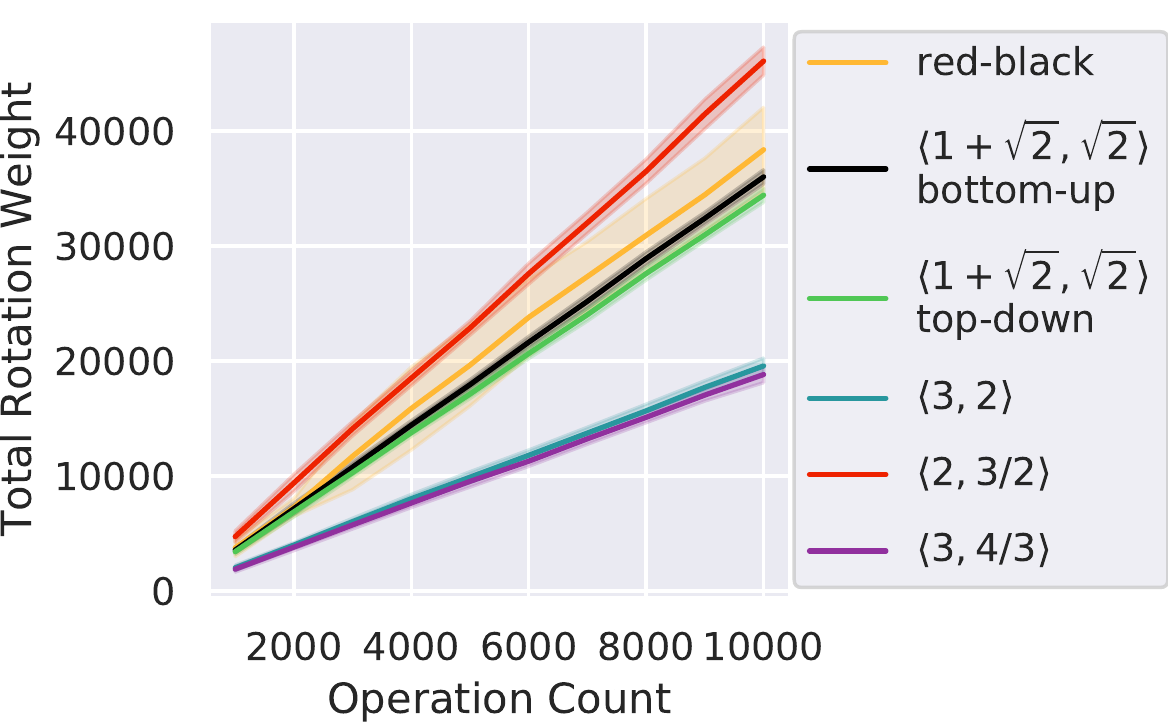}
  \caption{Total node weight of rotated nodes (on the $y$ axis) after a number of operations (on the $x$ axis).}
  \label{plt:rotation-size}

  \end{subfigure}
  \caption{Rotation count and rotated node weight for several kinds of trees of size $10^6$ after various numbers of operations. Solid lines report the mean value, while shaded areas indicate the standard deviation.}
  \label{plt:rotations}
\end{figure*}

For the final evaluation step, we consider that weight-balanced trees are often chosen because the total weight of the nodes rotated around can be theoretically bounded. This is useful if rotations around larger nodes are expensive, for example because of annotations that need to be repaired. We explore their behavior in this regard by creating random trees of size $10^6$, performing a number of operations (where every operation consists of one node removal and reinsertion with changed key) on them and counting how many rotations occurred, and what the total weight of the rotated nodes is. Figure~\ref{plt:rotation-count} shows how the number of rotations increases with increasing number of operations, Figure~\ref{plt:rotation-size} shows the same for the total weight of the rotated nodes. Note that we excluded the $\langle 3/2, 5/4 \rangle$ variant here, since $\Delta = 3/2$ is such a strong balancing requirement that the number of rotations is about $20$ times larger than for all the other variants, thus including it would have distorted the plot. The most striking point is that both numbers are significantly smaller for the variants with $\Delta = 3$, usually roughly half the number of rotations (resp.\ total rotation weight) than for the other variants or the red-black tree. The less strict balancing requirement apparently drastically reduces the number of necessary rotations. Whether one uses top-down or bottom-up balancing does not seem to make a serious difference.

It is also notable that the red-black tree, although not possessing a similar theoretical guarantee, does not perform significantly worse in terms of rotation count or weight  than the weight-balanced trees with $\Delta < 3$, although its total rotation weight has a much larger standard deviation. Consistent with the finding for $\Delta = 3$, the weight-balanced tree with $\Delta = 2$ performs the worst in terms of rotations.

\section {Conclusion}

In the paper on hand, we evaluated and engineered top-down weight-balanced trees. A rigorous evaluation has shown that using a top-down balancing approach instead of a bottom-up approach in fact leads to a significant performance increase, if one chooses the correct balancing parameters. The correct choice of balancing parameters can even make weight-balanced trees more performant than red-black trees, which is surprising considering the fact that red-black trees are used widely, while weight-balanced trees have received little attention in practice. However, the balancing parameters should be chosen with the intended use for the weight-balanced tree in mind. If little modification and a lot of searches are expected, we recommend using $\langle 2, 3/2 \rangle$ because of its superior average node depth. Even stronger balanced choices such as $\langle 3/2, 5/4 \rangle$ do not look advisable. One should also consider the expected distribution of nodes' keys. For strongly skewed distributions, as for example the \texttt{zipf} case, smaller $\Delta$ values such as $\langle 2, 3/2 \rangle$ tend to be advantageous. Also, for these distributions, weight-balanced trees should be chosen over red-black trees, as our analysis of average node depth has shown.

\balance
In case that the weight-balanced tree is annotated and rotations, especially around large nodes, are costly, using $\langle 3, 2 \rangle$ or even larger $\Delta$ values is likely to be the best of the evaluated choice. In fact, our benchmark of insert and deletion suggests that $\langle 3,2 \rangle$ and $\langle 3, 4/3\rangle$ are overall fairly performant choices, even if their average path lengths might be slightly inferior. It never seems to be a good choice to use the classic $\langle 1 + \sqrt{2}, \sqrt{2} \rangle$ variant. Summarizing these recommendations, it is surprising that empirically, many times the best choice for balancing parameters are parameters for which the theoretical guarantees do not hold, especially in the top-down rebalancing case. Of course, these parameters are only a viable choice if one does not have to worry about artificially crafted adversarial instances.

In the future, it would be interesting to determine the space of feasible balancing parameters for top-down weight-balanced trees similar to how Hirai and Yamamoto have done for bottom-up weight-balanced trees. 


\bibliography{bibliography}

\pagebreak
\appendix
\clearpage
\onecolumn
\nobalance

\section{Code and Data Publication}
\label{sec:appendix-data}

All evaluated trees as well as all benchmarking code is implemented in C++17. We publish the code (including all benchmarking code) at

\begin{center}
  \url{https://github.com/tinloaf/ygg/}
\end{center}

\noindent Note that this is an ongoing project subject to changes. The exact code revision used in this paper can be accessed at

\begin{center}
  \url{https://github.com/tinloaf/ygg/releases/tag/version_used_for_alenex20}
\end{center}

\noindent The code includes a file \texttt{README.md} with build instructions. After building, the directory ``benchmark'' contains all binaries necessary to reproduce our benchmarks. The file \texttt{BENCHMARKING.md} contains instructions on how to run the benchmarks.

We also publish all raw results we obtained from the benchmarks in a separate data publication~\cite{data-publication}. This publication can be accessed at

\begin{center}
  \url{https://publikationen.bibliothek.kit.edu/1000098852}
\end{center}

\noindent It also contains a detailled description of the data format output by the various benchmarking tools.

\clearpage

\section{Omitted Benchmark Plots}%
\label{app:omitted-benchmark}%

\begin{figure*}[h]
  \centering
  \begin{subfigure}[t]{0.39\textwidth}
    \includegraphics[height=0.25\textheight]{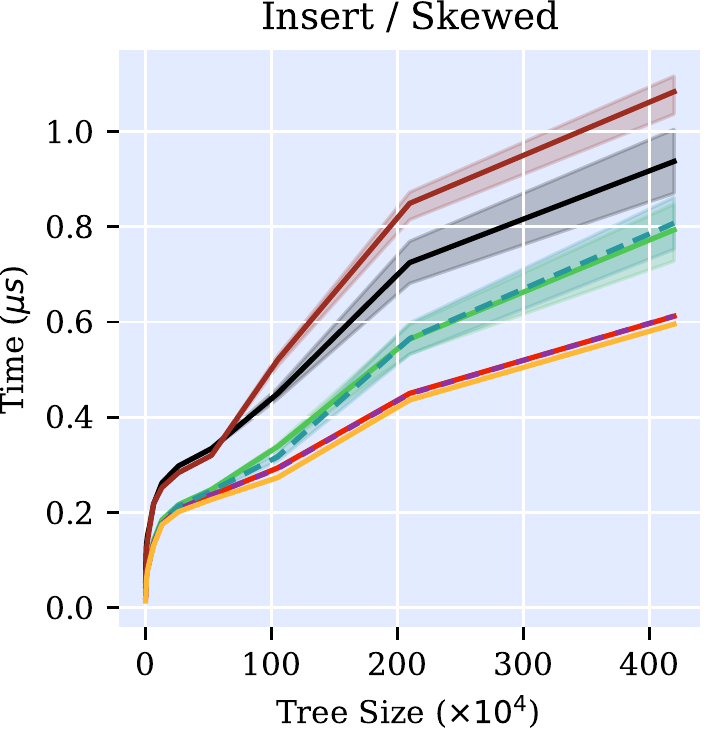}
    \caption{Nodes' keys chosen as for the \texttt{skewed} case.}%
    \label{plt:insert-time-skewed}
  \end{subfigure}
  \quad
  \begin{subfigure}[t]{0.57\textwidth}
    \includegraphics[height=0.25\textheight]{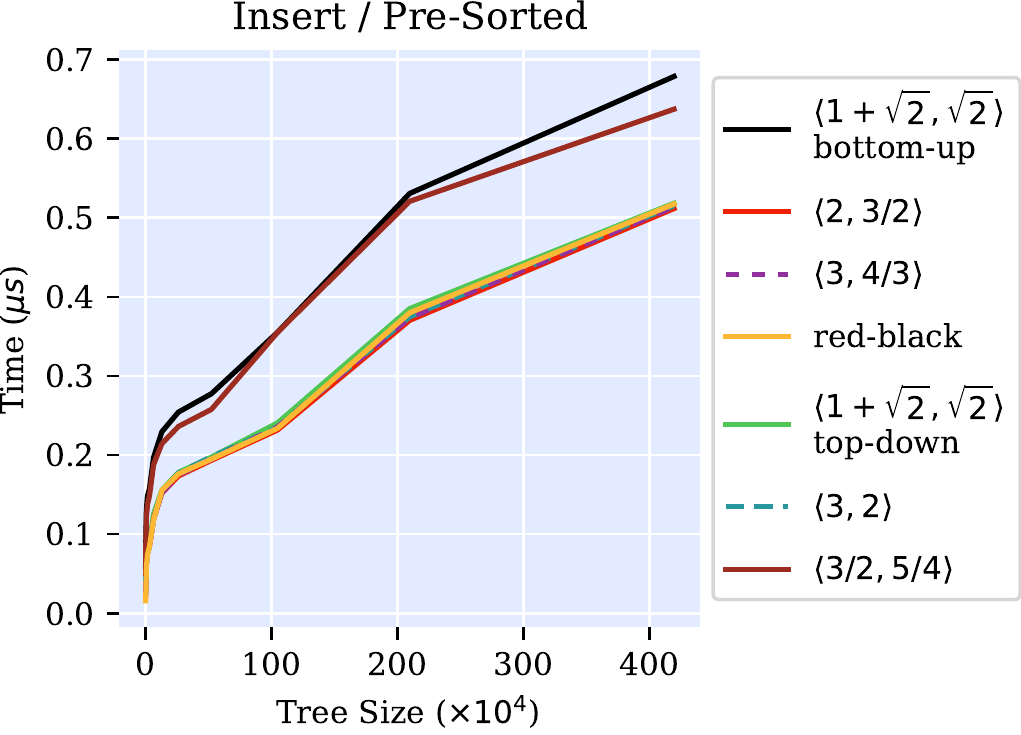}
    \caption{Nodes' keys chosen as for the \texttt{pre-sorted} case.}%
    \label{plt:insert-time-presorted}
  \end{subfigure}
  \caption{Times to insert $5\%$ new nodes into trees of various sizes. The $x$ axis specifies the size of the base tree. The $y$ axis reports the time needed for a single insertion in microseconds. Shaded areas indicate standard deviation.}
\end{figure*}%
\begin{figure*}[h]
  \centering

  \begin{subfigure}[t]{0.39\textwidth}
    \includegraphics[height=0.25\textheight]{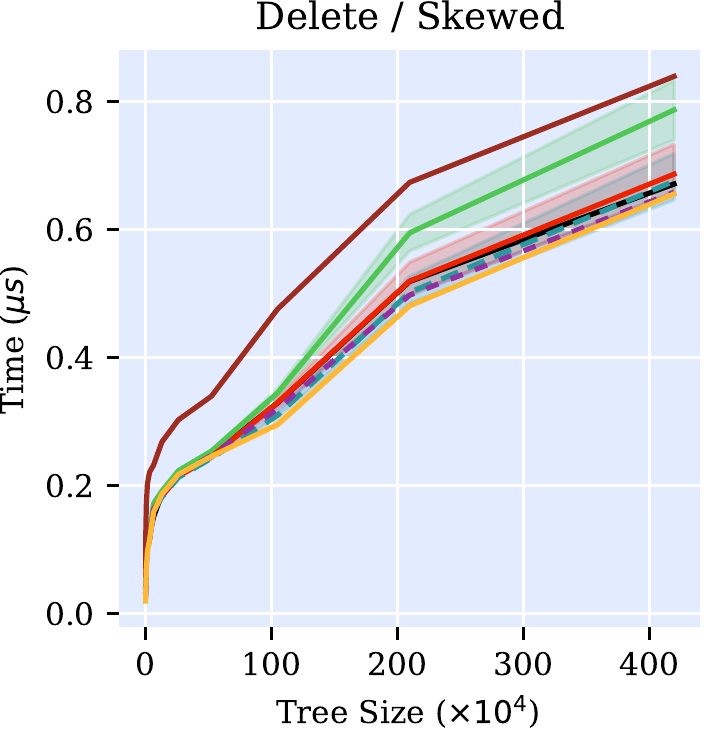}
    \caption{Nodes' keys chosen as for the \texttt{skewed} case.}%
    \label{plt:erase-time-skewed}
  \end{subfigure}
  \quad
  \begin{subfigure}[t]{0.57\textwidth}
    \includegraphics[height=0.25\textheight]{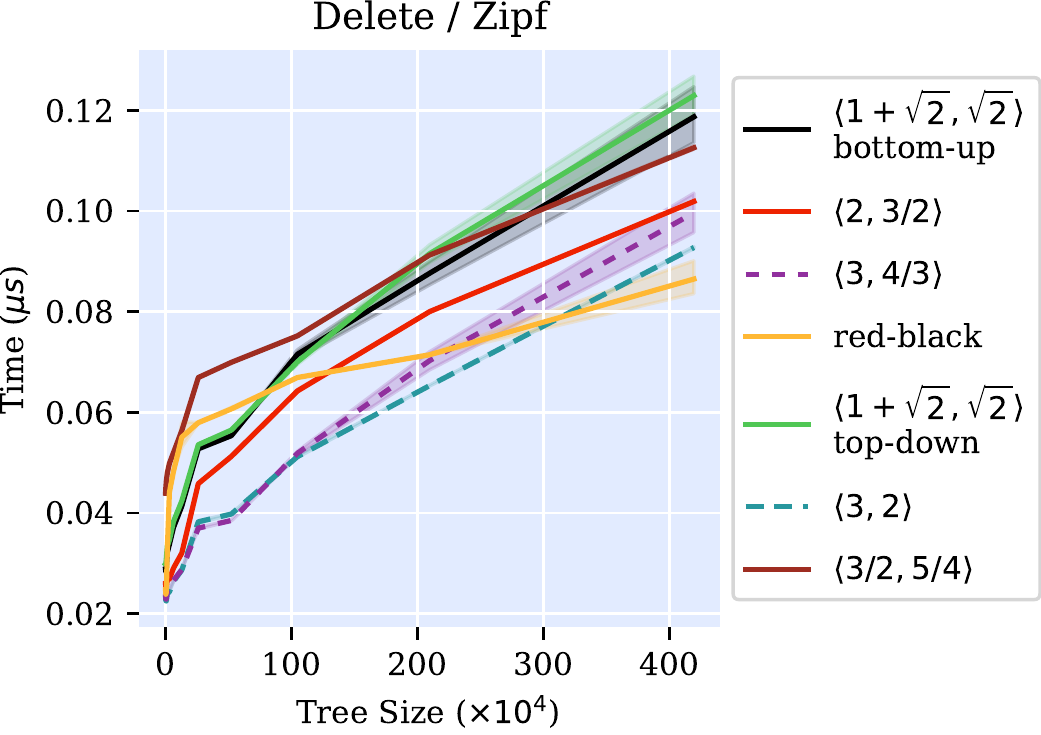}
    \caption{Nodes' keys chosen as for the \texttt{zipf} case.}%
    \label{plt:erase-time-zipf}
  \end{subfigure}
  \caption{Times to delete $5\%$ nodes from trees of various sizes. The $x$ axis specifies the size of the base tree. The $y$ axis reports the time needed for a single deletion in microseconds. Shaded areas indicate standard deviation.}
  \label{plt:erase-zipf-skewed}
\end{figure*}

\clearpage
\twocolumn

\begin{figure}[p]
  \centering
    \includegraphics[width=0.95\textwidth]{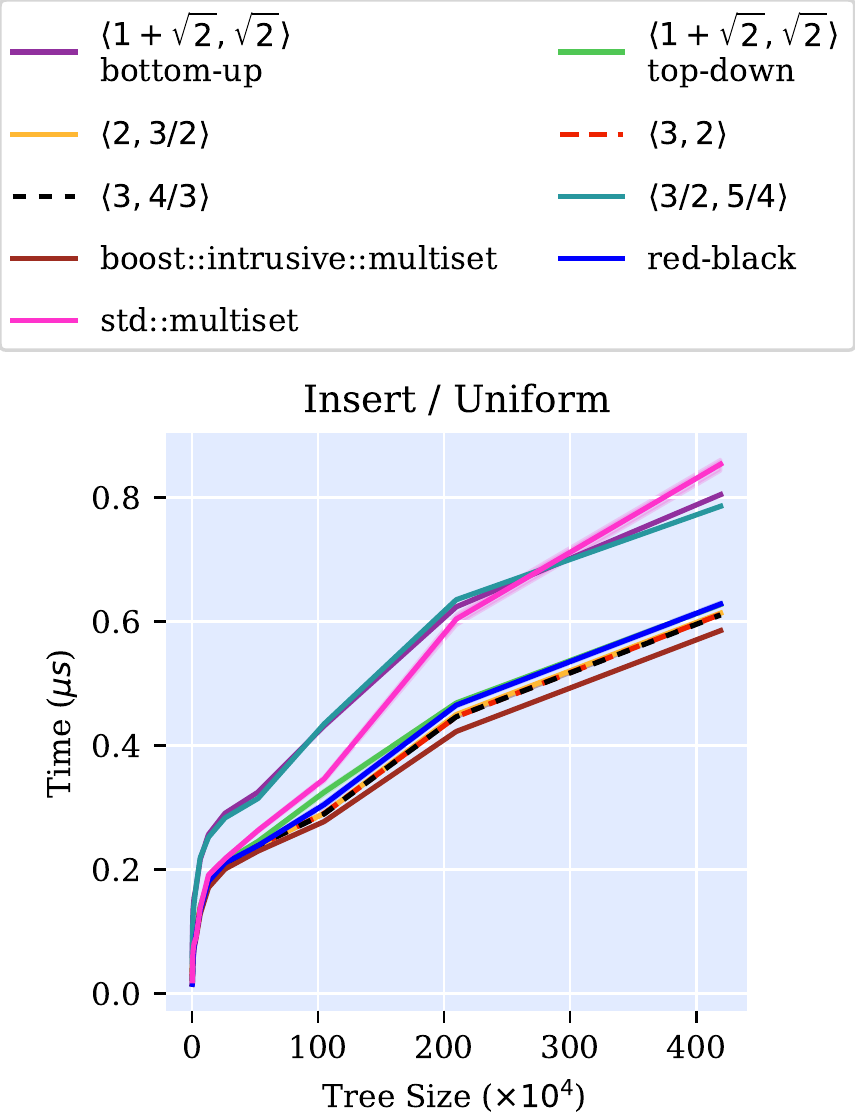}
  \caption{Times to insert $5\%$ new nodes into trees of various sizes. The $x$ axis specifies the size of the base tree. The $y$ axis reports the time needed for a single insertion in microseconds. Shaded areas indicate standard deviation. Node keys are chosen uniformly at random. Note that this plot includes \texttt{std::multiset} and \texttt{boost::intrusive::multiset}.}
    \label{plt:comparison-stdset}
\end{figure}

\begin{figure}[p]
  \centering
    \includegraphics[width=0.95\textwidth]{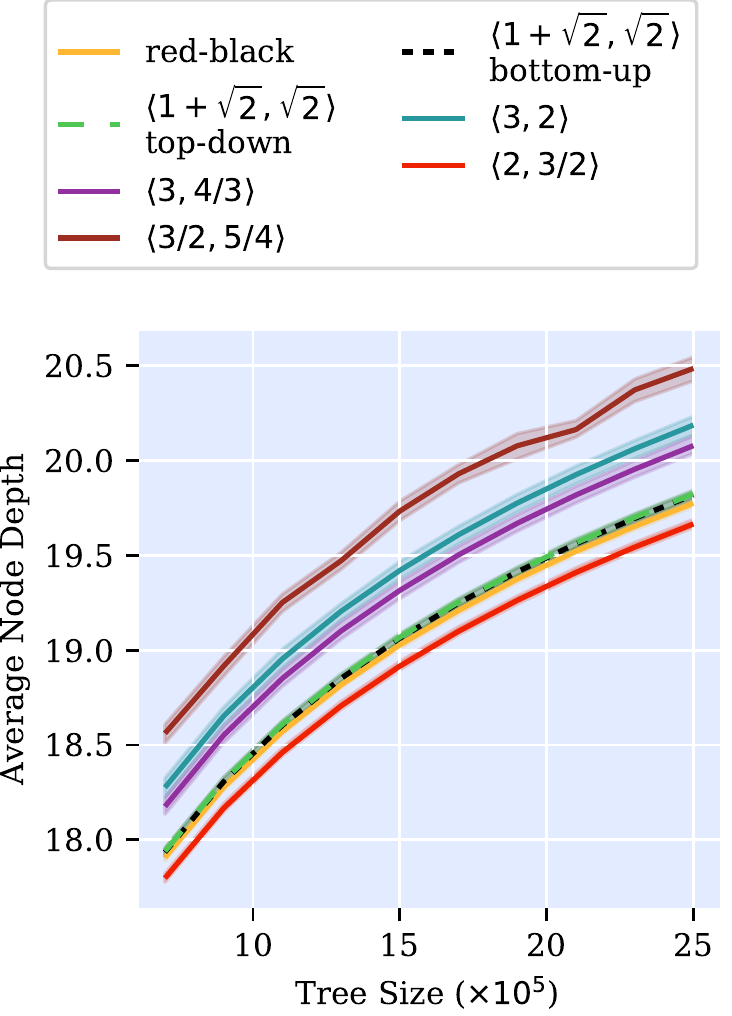}
  \caption{Average node depth for trees of various sizes, with keys chosen for the \texttt{skewed} case. The $x$ axis specifies the size of the tree, the $y$ axis the average node depth. All nodes in every tree were randomly generated, removed once, had their key changed, and were reinserted. The solid lines indicate average values, the shaded areas the standard deviation.}
    \label{plt:average-depth-skewed}
\end{figure}



\end{document}